# A single layer artificial neural network with engineered bacteria

Kathakali Sarkar[1,2], Deepro Bonnerjee[1,2], Sangram Bagh[1,*]

[1]Biophysics and Structural Genomics Division, Saha Institute of Nuclear Physics, Homi Bhabha National Institute (HBNI), Block A/F, Sector-I, Bidhannagar, Kolkata 700064 INDIA

[2]These authors contributed equally to the work.

* Corresponding author E-mail: sangram.bagh@saha.ac.in

**Abstract**: The abstract mathematical rules of artificial neural network (ANN) are implemented through computation using electronic computers, photonics and in-vitro DNA computation. Here we demonstrate the physical realization of ANN in living bacterial cells. We created a single layer ANN using engineered bacteria, where a single bacterium works as an artificial neuron and demonstrated a 2-to-4 decoder and a 1-to-2 de-multiplexer for processing chemical signals. The inputs were extracellular chemical signals, which linearly combined and got processed through a non-linear log-sigmoid activation function to produce fluorescent protein outputs. The activation function was generated by synthetic genetic circuits, and for each artificial neuron, the weight and bias values were adjusted manually by engineering the molecular interactions within the bacterial neuron to represent a specific logical function. The artificial bacterial neurons were connected as ANN architectures to implement a 2-to-4 chemical decoder and a 1-to-2 chemical de-multiplexer. To our knowledge, this is the first ANN created by artificial bacterial neurons. Thus, it may open up a new direction in ANN research, where engineered biological cells can be used as ANN enabled hardware.

**1. Introduction**

Artificial neural networks (ANNs) and its applications in various fields including robotics, speech recognition, complex pattern recognition, machine vision, autonomous vehicles, and many other domains inspire many technologists to speculate towards an 'era of technological singularity' [1]. Though such vision is debatable, ANNs' applications in almost all engineering and scientific disciplines including biomedical engineering and bigdata biological research are increasing at an exponential pace [1].

The physical realization of ANN is performed in conventional von Neumann computers, electronic machines, photonics, and in-vitro DNA computation [2,3], whose physical mechanism of operation is strikingly different from the biological neurons and their networks in the brain, which is the major inspiration for ANN [4]. Though, our understanding of neural network in the brain is far from complete, our understanding of ANN is from the first principle. This may allow creating the ANN at the cellular level by engineering its biochemical processes. Few computational studies tried to explain genetic networks from an ANN perspective [5,6]. However, as per our knowledge, the physical realization of ANN has not been done at the cellular level.

Here, we created single layer ANN architectures using engineered bacteria, where each bacterium works as an artificial neuron and we named it as 'bactoneuron' (BNeu). We demonstrated a 2-to-4 decoder and 2-to-1 demultiplexer, which decode input chemical signals, by using single layer neural networks made from bactoneurons.

## 2. Results and Discussion

### 2.1 Mapping of a 2-to-4 decoder using an ANN consisting of bactoneurons.

A $N:2^N$ decoder converts N bit binary coded inputs into $2^N$ coded outputs in a one-to-one mapping fashion. Therefore, in a 2-to-4 decoder, there are 2 inputs and $2^2$ (4) outputs. For each combination of 2 inputs (total 4 possible combinations), there would be one distinct output to map that specific input combination. The electronic design with the truth table for a 2-to-4 decoder is shown in Figure 1a. The 2-to-4 decoder in the ANN model has been shown in Figure 1b, where $X_1$ and $X_2$ are the inputs, $w_{ji}(s)$ are the corresponding weights, $y_i(s)$ are the summation function, and $B_i(s)$ are the bias. The linearly combined summation function, $y_i(s)$ goes thought a non-linear activation function to produce the output $O_i(s)$. Depending on the sign and numerical values of weight and bias, the four artificial neurons represent four different functions (Fig. 1b).

We mapped this abstract ANN model into an engineered cellular model (Figure 1c), where engineered bacterial cells worked as artificial neurons (bactoneurons). In this bactoneuron model, we replaced the abstract inputs ($X_1$ and $X_2$) to two environmental chemical signals, anhydrotetracyline (aTc) and Isopropyl β- D-1-thiogalactopyranoside (IPTG). The abstract outputs ($O_1$, $O_2$, $O_3$, and $O_4$) were changed to four fluorescent proteins, mKO2, E2-Crimson, mTFP1, and mVenus respectively. The bactoneurons combined the inputs and executed appropriate log-sigmoid activation functions through synthetic genetic circuits [7], (Figure 2). Synthetic genetic circuits are human-designed molecular genetic constructs, which follow engineering design principles and work inside living cells [7]. The activation function generated by the synthetic gene circuits, for each artificial neuron, represented a specific logic function (Figure 1b and 1c).

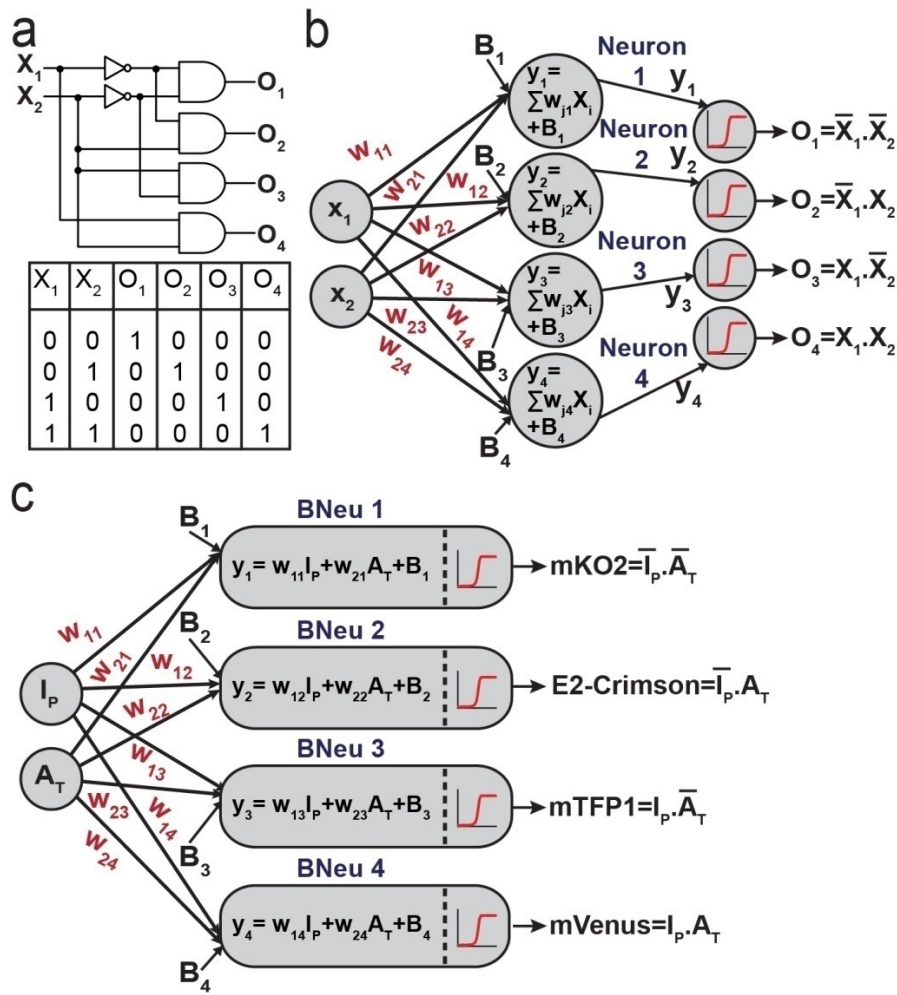

**Figure 1: a)** Electronic circuit diagram and truth table of a 2-to-4 decoder. **b)** ANN version of the 2-to-4 decoder. Each artificial neuron (1-4) receives weighted inputs $X_1$ & $X_2$, which are linearly combined into the individual neuron's summation function. The corresponding weights of the inputs ($w_{11}$, $w_{21}$, $w_{12}$, $w_{22}$, $w_{13}$, $w_{23}$, $w_{14}$, $w_{24}$) and the biases ($B_1$, $B_2$, $B_3$ and $B_4$) for each neuron are also shown. The linearly combined summation functions process through a non-linear activation function. Depending on the sign and values of the weight and bias, the activation functions represent four logical functions; a NOT $X_1$

AND NOT $X_2$ (Neuron 1), a $X_2$ N-IMPLY $X_1$ (Neuron 2), a $X_1$ N-IMPLY $X_2$ (Neuron 3) and a $X_1$ AND $X_2$ (Neuron 4). The ANN produces four outputs, namely, $O_1$, $O_2$, $O_3$ and $O_4$. **c)** Mapping of ANN with proposed artificial bacterial neurons; 'bactoneurons'. BNeu 1, BNeu 2, BNeu 3 and BNeu 4 represent four 'bactoneurons', which replace $X_1$ and $X_2$ with two chemical inputs, IPTG ($I_p$) and aTc ($A_T$) respectively and replaced outputs $O_1$, $O_2$, $O_3$ and $O_4$ with four fluorescent proteins mKO2, E2-Crimson, mTFP1 and mVenus respectively. The linear combinations of weighted inputs and non-linear activation function are performed appropriately by each bactoneuron.

**2.2 Engineering single bactoneurons**

We first engineered a single bacterium to function as a single artificial neuron, which performs a basic Boolean function through synthetic gene circuits. Those bactoneurons performed AND, NOR and two different N-IMPLY activation functions, which were implemented by specific designed genetic circuits (Figure 2). The genetic circuits were based on designed molecular interactions and feedbacks, such that it process extracellular chemical signals, IPTG and aTc in desired ways (NOR, N-IMPLY and AND).

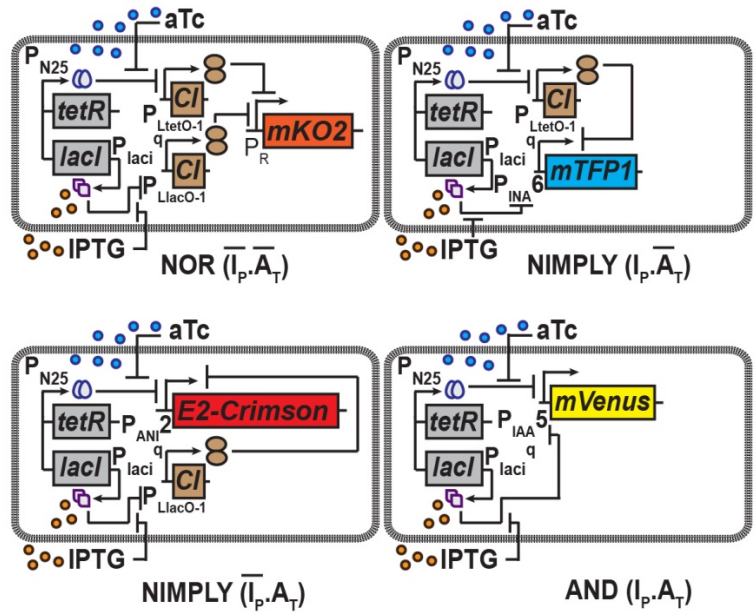

**Figure 2:** Genetic circuit design for four bactoneurons (BNeus). Promoter $P_{INA}6$, $P_{ANI}2$ and $P_{IAA}5$ are synthetic promoters designed to obtain desired function (promoter $P_{INA}6$ for IPTG N-IMPLY aTc,

promoter P$_{ANI}$2 for aTc N-IMPLY IPTG and promoter P$_{IAA}$5 for IPTG AND aTc) and carrying binding sites for TetR, LacI or CI transcription factors or its mutated variant. Every genes and promoters are instroduced in the cell through plasmids, except, LacI and TetR, which are expressed from *E. coli* DH5αZ1 genome by two different constitutive promoters as shown.

The genetic circuits combined the chemical signals and processed them through a log-sigmoid function (equation 1).

$$Output(y) = \frac{1}{(1+e^{-(A_T.wA+I_P.wI+B)})} \ldots\ldots \text{Equation (1)}$$

where,

A$_T$ represents aTc concentration,

w$_A$ represents the weight of input aTc for the neuron,

I$_P$ represents IPTG concentration,

w$_I$ represents the weight of input IPTG for the neuron,

B represents the Bias for the neuron.

The designs were based on hybrid promoters and their programmed interactions with various transcription factors. We created a library of constructs for each Boolean function and characterized for dynamic range and ultrasensitivity. The best candidates from the library have been chosen for further experimentations.

Figure 3 shows the realization of an AND bactoneuron. The AND bactoneuron (Figure 3a) was implemented through a synthetic genetic circuit (Figure 3b), which consists of a novel hybrid promoter (Figure 3c). The hybrid promoter binds transcription factors tetR and lacI, which can hinder the expression of EGFP from the same promoter in *E. coli* strain DH5αZ1, chassis for the bactoneuron, which constitutively expresses tetR and lacI [8]. Only in the simultaneous presence of IPTG and aTc, which bind with lacI and tetR respectively and change their conformation such that they cannot bind to the promoter anymore, the promoter is free to recruit RNA polymerase for EGFP expression. In order to achieve the AND bactoneuron, we designed, constructed and optimized eight different hybrid promoters (Figure 3c) in five consecutive sets (Figure 3d and Figure 3e). In each step the weight values were adjusted towards the desired behavior.

First, we characterized the behavior of an initial set of AND genetic devices containing the hybrid promoter $P_{IAA}1$, by measuring the EGFP expression at various combinations of 'zero' and 'saturated' concentration of IPTG and aTc (Figure 3d, column graph for $P_{IAA}1$). Next, we measured the dose response (data not shown) of the EGFP expression as a function of IPTG and aTc, by varying concentration of the one chemical, while keeping the other at constant saturated concentration. Further, the dose-response behaviors were fitted with a log-sigmoid function (equation 1), which is a conventional function for the artificial neuron characterization in ANN [4]. Here, for the linear combinations of two input signals, aTc and IPTG, we converted the concentration range of each input chemical from '0' to '1', where 0 signifies zero concentration and 1 signifies the saturating concentration of the chemical.

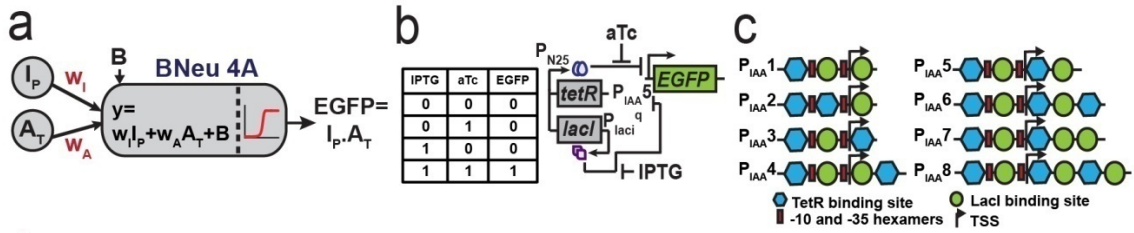
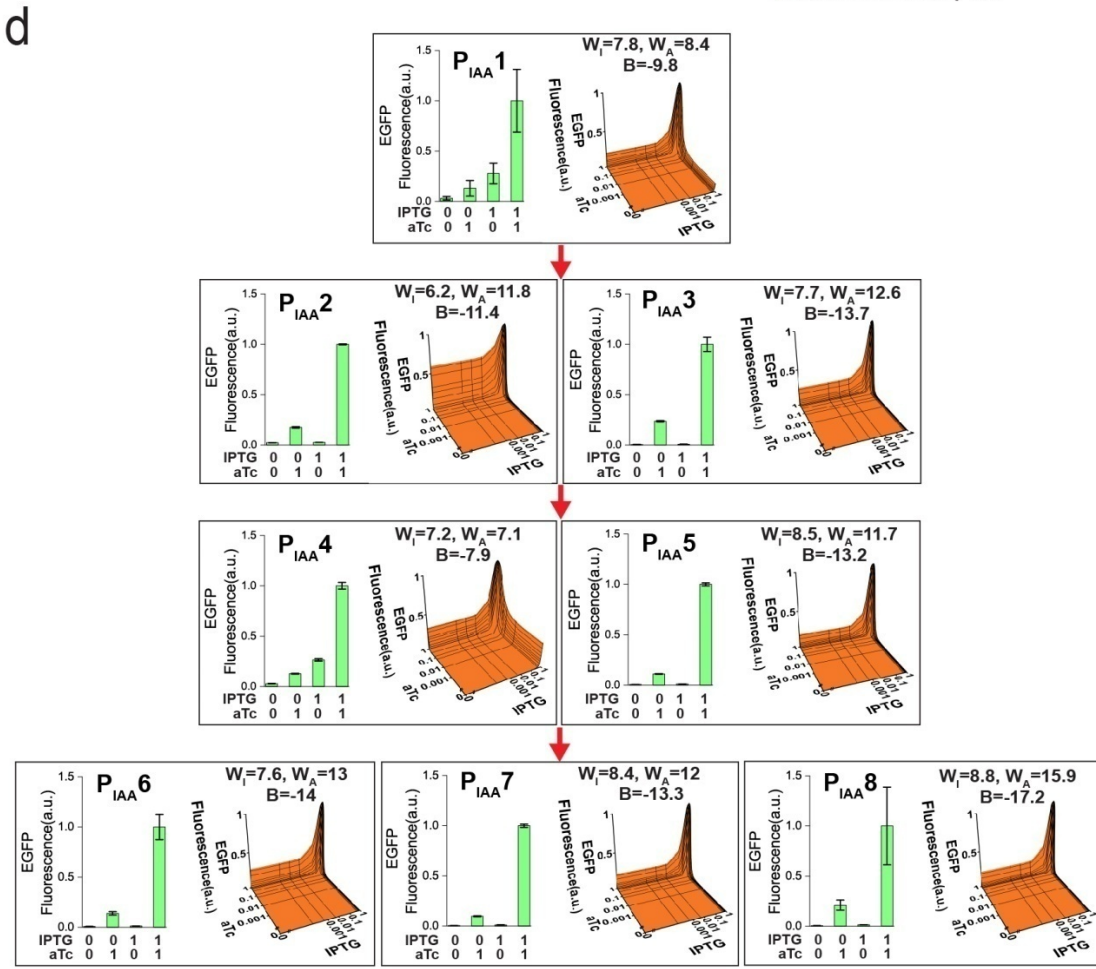
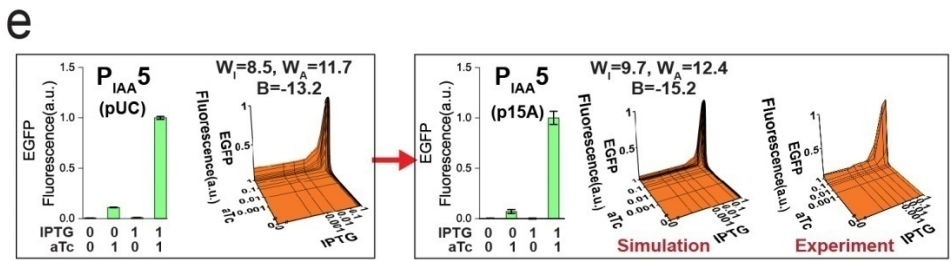

**Figure 3:** Characterization, weight and bias optimization of bactoneurons performing AND logic operation. **a)** Neural architecture of the bactoneuron BNeu 4A where $w_I$ and $w_A$ are weights of inputs IPTG ($I_P$) and aTc ($A_T$) and B is the bias. Summation function y is built from the weights and bias which gives rise to the activation function (showed by sigmoidal graph symbol) of the bacterial neuron BNeu 4A. BNeu 4A is same as BNeu 4 in figure 1c except the output gene is replaced to EGFP. Activation function generates output EGFP and EGFP production follows IPTG AND aTc logic behavior.
**b)** Truth table and biological circuit design of the IPTG AND aTc activation function. $P_{IAA}1-8$ are synthetic promoters designed for AND function and regulated by IPTG and aTc as they contain binding cites for LacI and TetR proteins. **c)** Schematic representation of promoter maps of $P_{IAA}1-8$ promoters. Positions of -10 and -35 hexamers, transcription start site and LacI & TetR binding sites are depicted in individual promoter map. **d)** Experimental characterization of truth table behavior (column graph) and simulated behavior (3D plot) of bacterial neuron carrying each $P_{IAA}$ promoter are shown. Values of weights and bias for individual system obtained from IPTG and aTc dose response experiments are also shown. Weights and bias were adjusted in multiple steps (shown by arrow) by changing the number and relative positions of the transcription factor binding sites in $P_{IAA}1$ promoter which gave rise to another two promoters: $P_{IAA}2$ and 3. Weight and bias values for $P_{IAA}3$ promoter suggests betterment of the system. Same strategy was followed again to develop promoter $P_{IAA}4-5$ followed by promoter $P_{IAA}6-8$. $P_{IAA}5$ showed the optimum IPTG AND aTc logic behavior. **e)** Weights and bias of the bacterial neuron carrying promoter $P_{IAA}5$ was again adjusted by changing the copy number of the plasmid carrying $P_{IAA}5$-EGFP gene cassette from high copy origin (pUC) to low copy origin (p15A). Simulated output behavior of this new system was compared with the experimental observation.

Any concentration higher than saturation concentration was treated as 1. The fitting parameters gave the estimation of the weight 'w' and the bias 'B' for each input. The 'weight' '$w_i$' and bias 'B' within a log-sigmoid function are shown in case of an IPTG, aTc AND function (Figure 3d, right panels, 3D plot for each promoter). Based on the estimated parameters, we performed a simulation for each AND bactoneuron (Figure 3d, right panels, 3D plot for each promoter). It is clear from the Figure 3d that the initial set of AND gates showed poor behavior. Next, we adjusted the weight and bias of the system manually by creating new promoters, where the number and positions of the operating sites for both TetR and LacI were varied. We performed this adjustment process a few times. Clearly, the weight 'w' of a bactoneuron is a strong function of types and degree of molecular interactions, as evident from the The fact that the weights of the initial AND bactoneuron can be adjusted to a new one with a new hybrid promoter of different molecular properties (Figure 3d). The circuit $P_{IAA}5$, with high weights for both $w_I$, $w_A$, which signifies the sharper transition from OFF to ON state in response, was taken for further weight adjustment by altering the relative numbers of transcription factors and copy numbers of the promoters by changing the copy number of the plasmids (Figure 3e). We found better weight values and less leakage. This bactoneron was selected as the final neuron for the AND function. We, further experimentally tested the AND neuron by simultaneously changing the concentration of the IPTG and aTc (Figure 3e). The results showed a close topological match with the simulation (Figure 3e).

Similarly, we created and characterized, fitted, simulated and re-tested other neurons for NOR (Figure 4a,b and c) and two different N-IMPLY functions (Figure 4d,e,f and Figure 4g,h,i). In all those bactoneurons, we used EGFP as an output signal. Figure 4c, 4f, and 4i showed that simulation and experiment are matched well for the NOR and two N-IMPLY bactoneurons, respectively. All the weight and bias values are tabulated in Table 1.

**Table 1:** Weights for IPTG ($w_I$), aTc ($w_A$) and bias for bactoneurons.

| Bacterial Neuron | $w_I$ | $w_A$ | Bias |
|---|---|---|---|
| BNeu 1A | -10 | -11.7 | 3.6 |
| BNeu 2A | -10.9 | 10.6 | -4.8 |
| BNeu 3A | 10.8 ($w_{IPTG1}$) | -14.9 ($w_{aTc1}$) | -3.4 |
| BNeu 4A_PIAA1_pUC | 7.8 | 8.4 | -9.8 |
| BNeu 4A_PIAA2_pUC | 6.2 | 11.8 | -11.4 |
| BNeu 4A_PIAA3_pUC | 7.7 | 12.6 | -13.7 |
| BNeu 4A_PIAA4_pUC | 7.2 | 7.1 | -7.9 |
| BNeu 4A_PIAA5_pUC | 8.5 | 11.7 | -13.2 |
| BNeu 4A_PIAA5_p15A | 9.7 | 12.4 | -15.2 |
| BNeu 4A_PIAA6_pUC | 7.6 | 13 | -14 |
| BNeu 4A_PIAA7_pUC | 8.4 | 12 | -13.3 |
| BNeu 4A_PIAA8_pUC | 8.8 | 15.9 | -17.2 |
| BNeu 1 | -10 ($w_{11}$) | -11.7 ($w_{21}$) | 3.6 |
| BNeu 2 | -10.9 ($w_{12}$) | 10.6 ($w_{22}$) | -4.8 |
| BNeu 3 | 10.8 ($w_{13}$) | -14.9 ($w_{23}$) | -3.4 |
| BNeu 4 | 9.7 ($w_{14}$) | 12.4 ($w_{24}$) | -15.2 |
| BNeu 4B | 9.7 ($w_{IPTG2}$) | 12.4 ($w_{aTc2}$) | -15.2 |

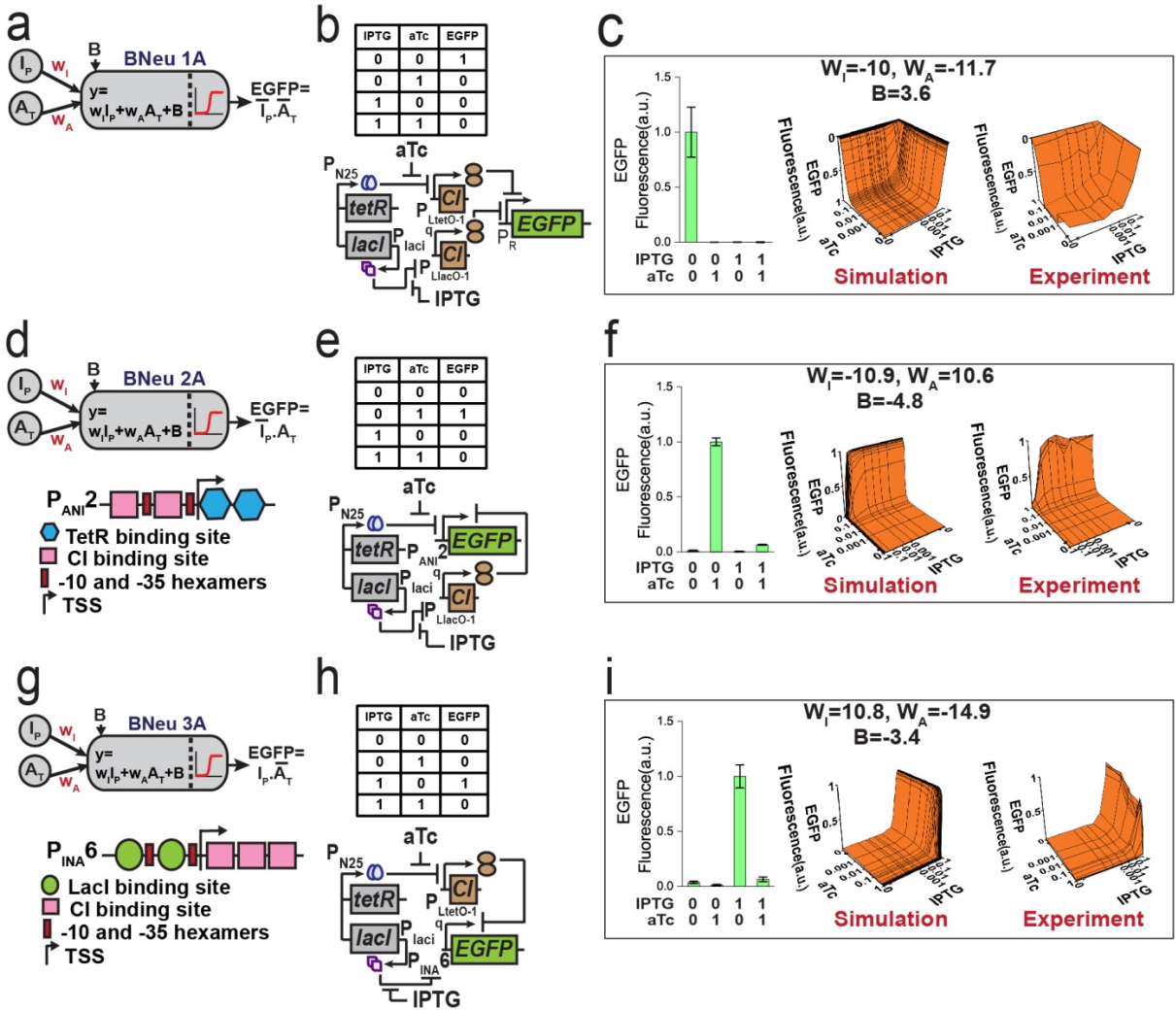

**Figure 4: a)** Neural architecture of bactoneuron BNeu1A performing NOR logic function. IPTG ($I_P$) & aTc ($A_T$) are the two inputs and $w_I$ & $w_A$ are their respective weights. **b)** Truth table and biological circuit design of (NOT IPTG) AND (NOT aTc) NOR. **c)** Truth table characterization, simulated and experimental output behavior of BNeu 1A. Values of Weights and bias are shown. **d)** Neural architecture of bactoneuron BNeu2A performing aTc N-IMPLY IPTG logic function. **e)** Truth table and genetic circuit design of aTc N-IMPLY IPTG. **f)** Truth table characterization, simulated and experimental output behavior of BNeu 2A. **g)** Neural architecture of bactoneuron BNeu 3A performing IPTG N-IMPLY aTc logic function. **h)** Truth table and genetic circuit design of IPTG N-IMPLY aTc. **i)** Truth table characterization, simulated and experimental behavior of BNeu 3A.

**2.3 ANN with bactoneurons gives a 2-to-4 decoder**

Next, we changed the EGFP output of the four individual bactoneurons (BNeu1, BNeu2, BNeu3, and BNeu4 ) to four distinct fluorescence proteins (mKO2, E2-Crimson, mTFP1, and mVenus) as shown in Figure 1c. The corresponding weight and bias values are shown in table 1. Those four bactoneurons in a mixed population should give the single layer ANN as shown in figure 1b and 1c. Those four artificial neurons were co-cultured and exposed to four various combinations of input chemical signals. The results are shown in Figure 5, which shows a good match with the expected truth table.

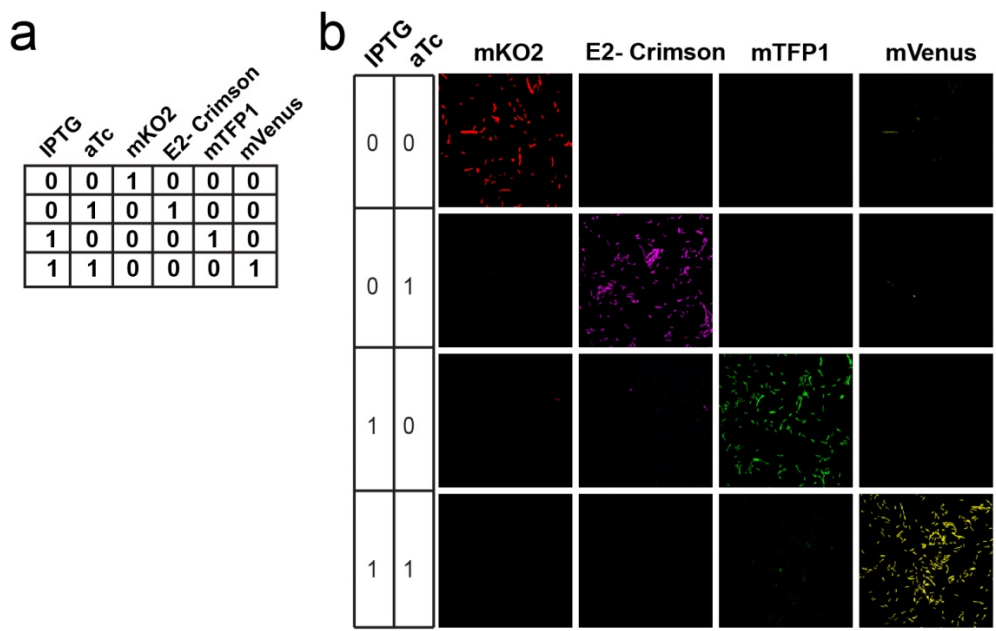

**Figure 5: a)** Truth table of the biological 2-to-4 decoder. **b)** Experimental behavior of the biological 2-to-4 decoder made from ANN consisting bactoneurons, studied with fluorescence microscope. Each row corresponds to each out of four 2-inputs logic states (IPTG-aTc) and four columns represent respective outputs (mVenus, mTFP1, E2-Crimson and mKO2) in terms of reporter genes expression coming from individual bactoneurons.

**2.4 1-to-2 De-multiplexer with bactoneuron based ANN**

One of the most important advantages of creating single bacterial neurons with the specific function was to create other functions with the same individual neurons just by mixing and matching through ANN architectures. To demonstrate that, we created one 1-to-2 De-multiplexer (Figure 6) by choosing a set of neurons from the bactoneurons library, we reported above (Table 1). Here, we changed the mTFP1 output of BNeu 3 with EGFP and mVenus output of BNeu 4 to E2-Crimson (Figure 6d). The two bactoneurons were co-cultured with appropriate chemical signals following the truth table (Figure 6b) and visualized the results in a fluorescent microscope (Figure 6e). The result matched with the truth table.

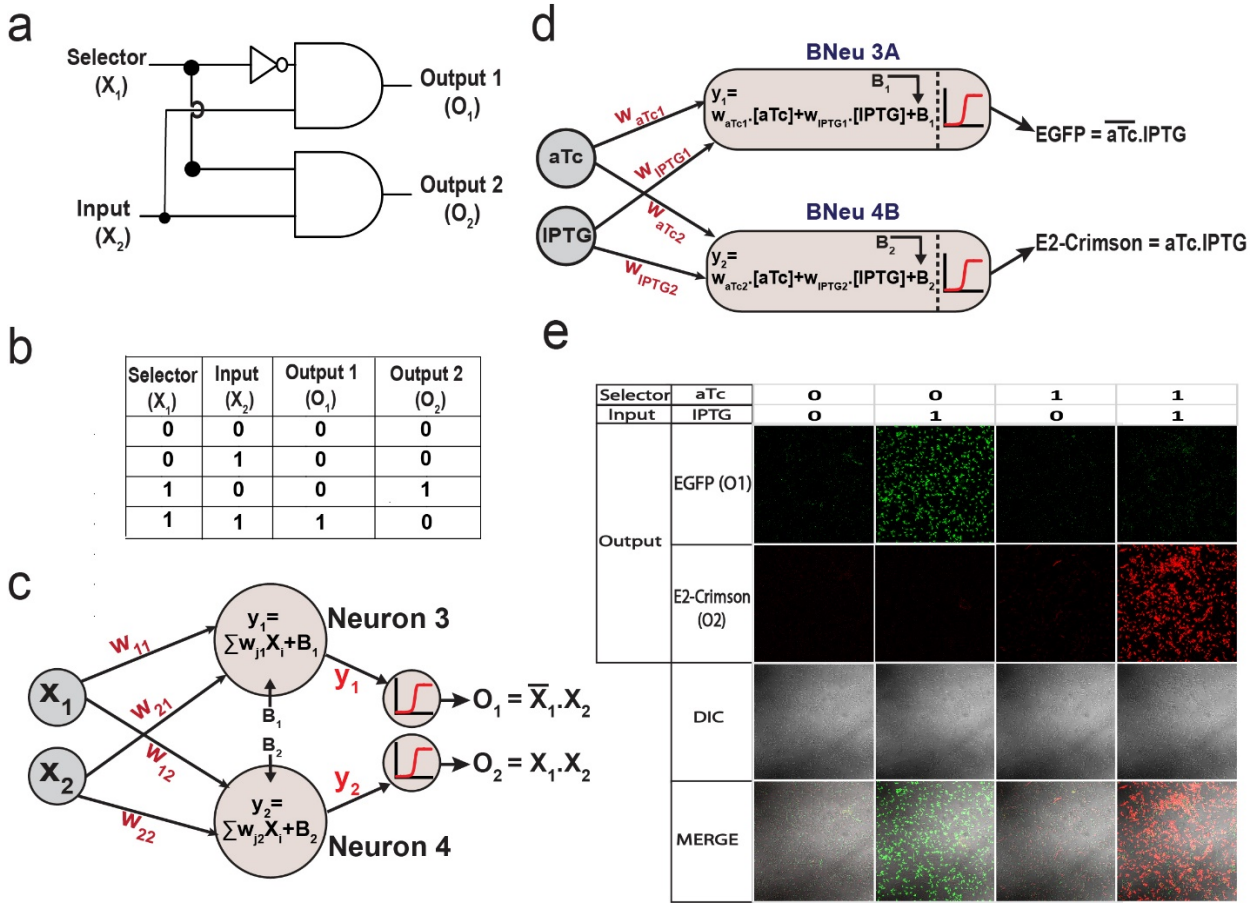

**Figure 6: a)** Electronic circuit diagram of a 1-2 Demultiplexer. **b)** Truth table of a 1-2 De-Multiplexer. **c)** ANN implementing for of a 1-2 De-Multiplexer. Each Artificial Neuron (Neuron 3 and 4) receives weighted inputs $X_1$ & $X_2$ that is fed to the individual neuron's summation function. The output from the summation function activates an activation function from which the artificial neuron's output is obtained. Depending on the values of the weights of the inputs ($w_{11}$, $w_{21}$, $w_{12}$, $w_{22}$) and the biases ($B_1$ & $B_2$) the artificial neurons produce outputs following the function $X_1$ N-IMPLY $X_2$ (Neuron 3) or $X_1$ AND $X_2$ (Neuron 4). **d)** Bactoneuron based realization of a 1-2 De-Multiplexer. BNeu 3A and BNeu 4B represent two engineered *E coli*, respectively implementing the functions of Neuron 3 and 4. The inputs $X_1$ & $X_2$ is served by two chemical small molecules aTc and IPTG respectively which are received by BNeu 3A and BNeu 4B through different weights. The weighted inputs are summed up by linear combination and processed through a log-sigmoid activation functions. Both the linear summation and non-linear activation functions are processed through the bactoneuron containing synthetic genetic circuits. Two fluorescent proteins EGFP and E2-Crimson respectively, serve as the Outputs $O_1$ & $O_2$. **d)** Both the bactoneurons are co-cultured with approriate chemical inputs. The fluorescence microscopy images show the proper functional behaviour of a 1-2 De-multiplexer.

## 3. Conclusions

Here we showed that ANN could be physically realized in engineering bacteria, where a single bacterium worked as an artificial neuron. The inputs were extracellular chemical signals, which linearly combined and transferred through a non-linear activation function to produce a fluorescent protein output. The log-sigmoid activation function was generated by synthetic genetic circuits and for each artificial neuron, the weight and bias values were adjusted manually by engineering the molecular interactions within the bacterial neuron to represent a specific logical function. Single bacterial neurons, representing a single basic logic function inspired to create more complex function by just allowing more neurons to connect through a single layer network. The artificial bacterial neurons were connected through a network to create a 2-to-4 chemical decoder and a 1-to-2 chemical de-multiplexer. To our knowledge, this is the first ANN created by engineered bacteria, which work as artificial neurons. We believe that this study shows a step towards using living bacteria as ANN abled hardware.

## 4. Materials & Methods

### 4.1. Bacterial strains and reagents

Chemically competent *Escherichia coli* DH5α strain was used for cloning and amplification of the genetic devices in plasmids and DH5αZ1 strain was used for the experimental characterization of the genetic devices. Working concentrations of the antibiotics in LB-Agar, Miller (Difco, Beckton Dickinson) plates as well as in LB broth, Miller (Difco, Beckton Dickinson) were: 100 μg/ml for Ampicillin (Himedia), 34 μg/ml for Chloramphenicol (Himedia) and 50 μg/ml for Kanamycin (Sigma Aldrich). IPTG and aTc were purchased from Abcam, Sigma Aldrich and Himedia. PCR amplification for target DNAs was performed by KOD Hot Start DNA polymerase (Merck Millipore), Pfu Turbo

Hotstart PCR Master Mix (Agilent Technologies) or Phusion High-Fidelity PCR Master Mix with HF Buffer (New England BioLabs). Restriction enzymes, T4 DNA ligase and DNA ladders were from New England BioLabs. Plasmid isolation, gel extraction and PCR purification kits were from QIAGEN.

**4.2. Promoters and genes, plasmids, RBS's & primers**

All primers and oligos were provided by Integrated DNA Technologies. All synthesized gene products were supplied by Invitrogen GeneArt Gene Synthesis service, Thermo Fischer. The promoters were obtained by either PCR from overlapping primer pairs or individually synthesized as gene products, being individually flanked by compatible primers sites. Primers were so designed to incorporate flanking Xho1 & ecoR1 sites or *Xho*I & *Kpn*I sites (including RBS sequence), as the case may be, to the PCR products which were digested accordingly for site specific cloning. The cloning backbones contained Ampicillin or Chloramphenicol resistance and pUC, ColE1 or p15A origins of Replication as the case may be. The vector backbones with appropriate cloning sites, EGFP reporter Gene, PltetO1 and $P_R$ promoters, ColE1, pUC or p15A origin were obtained from plasmids pTA1EGFP, pRA1EGFP, pTA2EGFP, or pRC3EGFP [9]. Wild type λ repressor CI gene was obtained from pTC3cI [9]. Frame shifted CI gene and $P_{LlacO-1}$ promoter were obtained from pLA2ScIfmTcIfm [9]. Weak RBSs were fused to the EGFP gene during its PCR amplification from plasmid pRA1EGFP and between flanking *ECoR*I and *Xba*I sites. The fluorescent reporter proteins, mVenus, mTFP1, E2-Crimson and mKO2 were obtained from commercial plasmids pmVenus-C1 (ClonTech), mTFP1-Pbad (Plasmid#54553, Addgene), pUCP20T-E2Crimson (Plasmid #78473, Addgene) and mKO2-pBAD (Plasmid#54555, Addgene) and cloned with flanking *Kpn*I and *Xba*I sites at the upstream and downstream. All cloned genes, promoters, RBS's in plasmid constructs were sequence verified by Eurofins Genomics India Pvt. Ltd., Bangalore, India.

**4.3 Bacterial cell culture for characterization of genetic constructs**

DH5αZ1 cells were transformed with appropriate Sequence verified plasmid constructs. Well-isolated single colonies were picked from LB agar-plates, inoculated to fresh LB-liquid media and grown overnight in presence of antibiotics. Next, the overnight culture was re-diluted 100 times in fresh LB media with antibiotics and with or without inducers IPTG or aTc as per the design of the gene circuit, and grown for 6hrs (for $P_{IAA}$ 1-4), 12 hrs (for $P_{IAA}$ 5-8) and 16 hrs for the rest at 37º C, ~ 250 rpm. Cells were then washed and re-suspended in phosphate buffered saline (PBS, pH 7.4) and EGFP fluorescence was measured. Concentration corresponding to input logic level "0" for any input inducer was 0 units. For input logic level "1", we considered inducers' concentration as: 10 mM IPTG and 200 ng/ml aTc.

**4.4 Measurement of fluorescence and optical density, Normalization and Scaling**

For fluorescence and optical density (OD) measurements, Synergy HTX Multi-Mode reader (Biotek Instruments, USA) was used. For this purpose, cells were diluted in PBS (pH 7.4) to reach around $OD_{600}$ 0.8, loaded onto 96-well multi-well plate (black, Greiner Bio-One) and both EGFP fluorescence with appropriate gain and $OD_{600}$ was measured. For EGFP fluorescence measurements, we used 485/20 nm excitation filter and 516/20 nm emission bandpass filter. At least 3 biological replicates had been considered for each condition to collect the fluorescence and OD data. The raw fluorescence values were divided by respective $OD_{600}$ values and thus normalized to the number of cells. Auto-fluorescence was measured as average normalized fluorescence of the untransformed DH5αZ1 set (no plasmid set) and subtracted from the normalized fluorescence value of the experimental set. The above normalization can be mathematically represented as follows: -

$$\text{Unscaled Normalized Fluorescence} = \left[\frac{(\text{Absolute Fluorescence value from experimental cell population})}{(\text{OD of experimental cell population})}\right] - \left[\frac{(\text{Absolute Fluorescence value from No plasmid cell population})}{(\text{OD of No plasmid cell population})}\right]$$

The values thus obtained were then scaled down between 0 and 1-considering the normalized Fluorescence value at the induction point of maximum expected Fluorescence to be 1- by dividing all other fluorescence values by this fluorescence value.

$$\text{Scaled Normalized Fluorescence} = \left[\frac{(\text{Unscaled Normalized Fluorescence at any induction point})}{(\text{Unscaled Normalized Fluorescence at induction point of maximum expected Fluorescence})}\right]$$

**4.5. Dose response experiments and 3D-experiments**

All dose response experiments were performed by varying one inducer across 10 or more concentration points while the other inducer was kept constant ("0" state or "1" as the case may be). For the 3D experiments, the corresponding two inducers of the relevant constructs were simultaneously varied across 10+ concentration points.

DH5αZ1 cells were transformed with the appropriate sequence-verified plasmid construct(s). Single colonies from LB-Agar plate with required antibiotics were picked and cultured for 8-10 hours in LB-media with corresponding antibiotics. They were then diluted 100 times in fresh LB media with antibiotics and inducers IPTG or aTc as the induction case may be and again grown for 10 h at 37º C, ~ 270 rpm. Cells were then sub-cultured with 100 times re-dilution into fresh LB-liquid media having the corresponding induction state as before. This sub-culture was grown for 6 hours at 37º C, ~ 270 rpm. Cells were then washed and re-suspended in phosphate buffered saline (PBS, pH 7.4) for taking fluorescence measurements.

**4.6. Data Analysis, Fitting, Mathematical Modelling and Simulation**

An artificial neuron can be mathematically described by equation 1 as shown in the results section:

Dose response experiments were performed by varying one inducer across 10 concentration points while the other inducer was kept constant ("0" state or "1" as the case may be). Since one of the inputs was kept constant, the product of its logical state and its weight, plus B gives a new constant as b as demonstrated by the following equation.

$$Output(y) = \frac{1}{(1+e^{-(X_n.W_n+b)})} \quad \text{..... Equation (2a)}$$

where,

    **Y** is the output signal from the artificial neuron,

    **$X_n$** corresponds to the magnitude of the varying input (n) to the artificial neuron,

    **$w_n$** corresponds to the Weight of the $n^{th}$ input to the artificial neuron,

    and,

**b** is given by:

$$b = X_m \cdot W_m + B \quad \ldots \text{Equation (2b)}$$

where,

**$X_m$** corresponds to the logical state of the constant input (m) to the artificial neuron (0 or 1),

**$W_m$** corresponds to the Weight of the $m^{th}$ input to the artificial neuron,

**B** corresponds to the Bias of the artificial neuron

The scaled output fluorescence values obtained from the dose response experiments were plotted against the varying inducer concentration and fitted against equation (2a). All data analysis and fitting were performed in OriginPro 2018 (OriginLab Corporation, USA) and were performed using built-in Levenberg Marquardt algorithm, a damped least squares (DLS) method. Based on the fitting curve thus obtained, the inducer concentration points were scaled such that the concentration value where scaled EGFP Normalized Fluorescence reaches saturation, was considered to be "1". The dose response data plot was then re-fitted with equation (2a) with the newly scaled input values. At this juncture, the parameter "$W_n$" of the fitting function (equation (2a)) gives the "weight" of the varying input in the summation function of the corresponding neuron. The b value obtained includes the Bias plus the product of the Input logic state of the second input and its weight as explained above. Upon similarly fitting the dose response of the neuron to the second input, the weight "W" and "b" for input2 is obtained. With the thus derived "$b_{input1}$" and "$b_{input2}$", the Bias "B" for the Summation function of the neuron is calculated. Subtracting the product of $W_{input-n}$ and the logic state of the same input (0 or 1) from the "b" value obtained from fitting the dose response of the other input gives us the Bias of the complete Summation Function. With the thus derived "$W_{input1}$" and "$W_{input2}$" and "B", the complete Summation function of the corresponding Neuron is accordingly built and is fed to the Activation Function given by equation 1. The simulations were performed by generating matrices of calculated normalized output fluorescence values against simultaneously varying concentrations of the corresponding two inputs across 50X50 points following the parameterized activation function.

**4.7 Microscopy**

DH5αZ1 cells were transformed with the appropriate sequence-verified plasmid Construct(s). Following an aforementioned 10+6 hour induction step, cells were washed thrice in PBS. Cell pellets were finally re-suspended in 300uL PBS and this re-suspension was used to prepare fresh slides. A Laser Scanning Microscope Zeiss LSM 710/ ConfoCor 3 operating on ZEN 2008 software was used for imaging. The cell suspension slides were subjected to excitation by appropriate laser channels (458 nm Ar Laser for mTFP1, 488 nm Ar Laser for EGFP, 514 nm Ar Laser for mVenus, 543 nm He-Ne Laser for mKO2 and 633 nm He-Ne Laser for E2-Crimson) and fluorescence emissions were captured through proper emission filters (BP484-504 nm for mTFP1, BP500-520 nm for EGFP, BP521-541 nm for mVenus, BP 561-591 nm for mKO2, BP641-670nm (2-to-4 decoder)/BP630-650 nm (1-to-2 demultiplexer) for E2-Crimson) with a 63x Oil immersion objective and were detected through a T-PMT.

**Acknowledgements**

This work was supported by SINP intramural funding, Department of Atomic Energy, Govt. of India, and partially funded by Ramanujan Fellowship of SB, Department of Science and Technology, Govt. of India.